\documentclass{emulateapj}

\slugcomment{to appear in The Astrophysical Journal (December 2006)}
\newcommand{\be}{\begin{equation}}
\newcommand{\ee}{\end{equation}}

\begin{document}

\title{On the Search For Transits of the Planets Orbiting Gl 876}
\author{P.D. Shankland\altaffilmark{1,2}, E.J. Rivera\altaffilmark{3},
G. Laughlin\altaffilmark{3}, D.L. Blank\altaffilmark{1}, A.
Price\altaffilmark{4}, B. Gary\altaffilmark{4}, R.
Bissinger\altaffilmark{4}, F. Ringwald\altaffilmark{5}, G.
White\altaffilmark{1}, G.W. Henry\altaffilmark{6}, P.
McGee\altaffilmark{7}, A.S. Wolf\altaffilmark{3}, B.
Carter\altaffilmark{8}, S. Lee\altaffilmark{9}, J.
Biggs\altaffilmark{10}, B. Monard\altaffilmark{11}, M.C.B.
Ashley\altaffilmark{12} }

\altaffiltext{1}{Centre for Astronomy, School of Mathematical \&
Physical Sciences, James Cook University, Townsville QLD 4811 AU}

\altaffiltext{2}{Current Address: U. S. Naval Observatory, 3450
Massachusetts Ave NW, Washington, D.C. USA 20392-5420
(paul.shankland@usno.navy.mil)}

\altaffiltext{3}{ University of California/Lick Observatory,
University of California at Santa Cruz, Santa Cruz, CA USA 95064}

\altaffiltext{4}{ American Association of Variable Star Observers,
Clinton B. Ford Astronomical Data \& Research Center, 25 Birch St.,
Cambridge MA USA 02138 (http://www.aavso.org)}

\altaffiltext{5}{ Department of Physics, California State
University, Fresno, 2345 E. San Ramon Ave., M/S MH37, Fresno, CA USA
93740}

\altaffiltext{6}{  Center of Excellence in Information Systems,
Tennessee State University, 3500 John A. Merritt Blvd., Box 9501,
Nashville, TN 37209 USA}

\altaffiltext{7}{ School of Chemistry \& Physics, University of
Adelaide, Adelaide SA 5005 AU}

\altaffiltext{8}{ Centre for Astronomy, Solar Radiation \& Climate
Physics, University of Southern Queensland, Towoomba 4350 AU}

\altaffiltext{9}{ Anglo-Australian Observatory, 167 Vimiera Rd,
Eastwood NSW 2122 AU}

\altaffiltext{10}{ Perth Observatory, Walnut Rd, Bickley WA 6076 AU}

\altaffiltext{11}{ Bronberg Observatory, Pretoria, South Africa}

\altaffiltext{12}{ School of Physics, University of New South Wales,
Sydney, NSW 2052 AU}

\begin{abstract}
We report the results of a globally coordinated photometric campaign
to search for transits by the $P\sim30 \, {\rm d}$ and $P\sim60 \,
{\rm d}$ outer planets of the 3-planet system orbiting the nearby
M-dwarf Gl 876. These two planets experience strong mutual
perturbations, which necessitate use of a dynamical (four-body)
model to compute transit ephemerides for the system. Our photometric
data have been collected from published archival sources, as well as
from our photometric campaigns that were targeted to specific
transit predictions. Our analysis indicates that transits by planet
``c'' ($P\sim30 \, {\rm d}$) do not currently occur, in concordance
with the best-fit $i=50^{\circ}$ co-planar configuration obtained by
dynamical fits to the most recent radial velocity data for the
system. Transits by planet ``b'' ($P\sim60 \, {\rm d}$) are not
entirely ruled out by our observations, but our data indicate that
it is very unlikely that they occur. Our experience with the Gl 876
system suggests that a distributed ground-based network of small
telescopes can be used to search for transits of very low mass
M-stars by terrestrial-sized planets.
\end{abstract}


\section{Introduction}
The Gl 876 planetary system ranks with the most remarkable
dicoveries that have emerged from the first decade of extrasolar
planet detections. As described by Rivera et al. (2005), Laughlin et
al. (2005), Butler et al. (2001) Marcy et al. (2001), Delfosse et
al. (1998) and Marcy et al. (1998), there are a number of reasons
why the Gl 876 planets are extraordinary. Planet ``d", with its
1.974 day orbit and $M \sin(i)=5.9 M_{\oplus}$ is the lowest-mass
exoplanet known to orbit a nearby main-sequence star (Rivera et al.
2005). Planets ``c" and ``b" are the only Jovian-mass companions
known to orbit an M-dwarf, and their clear participation in 2:1
mean-motion resonance gives important dynamical clues to the
formation and evolution of the system (e.g. Lee and Peale 2002).
Furthermore, the Gl 876 planets are the closest exoplanets that have
been reliably characterized. Their radial velocity amplitudes induce
an overall signal-to-noise of nearly 100, and they have have been
observed for well over a decade.

The detection of Gl 876 d by Rivera et al. (2005) underscored the
rapid development of the Doppler technique. The detection of planets
having only a few Earth masses is the culmination of advances
described by Butler et al. (1996), Marcy \& Butler (1998), Marcy,
Cochran, \& Mayor (2000), Marcy et al. (2005), and Lovis et al.
(2006). Nevertheless, while radial velocity precision has improved
to better than 1 m/s, Keplerian orbital fits to Doppler velocities
give estimates of $M_{\rm pl}\sin(i)$ rather than $M_{\rm pl}$. In
order to ascertain the true mass of a planet, one must obtain an
independent measure of the orbital inclination, $i$, which is most
easily measured when the planet is observed to transit across the
face of the parent star.

The now-celebrated companion to HD 209458 provided the first example
of a transiting extrasolar planet, and this body has generated an
incredible variety of observational and theoretical investigations.
HD 209458b was initially detected with the radial velocity method,
and was then discovered to transit on the basis of follow-up
photometric observations at the predicted transit times (Henry et
al. 1999, 2000; Charbonneau et al. 2000). HD 209458b's well-sampled
light curve, when combined with high-precision radial velocity data,
has permitted precise measurements of the planet's basic attributes
(e.g. Mazeh et al. 2000, Brown et al. 2001, Laughlin et al. 2005,
Winn \& Holman  2005).

To date, nine additional planets have been found to transit their
parent stars, and the varied uses to which these transits have been
put are described in the review article by Charbonneau et al.
(2006). Our goal in this paper is to ascertain whether the two outer
planets orbiting Gl 876 (IL Aqr, GJ 876, HIP 113020) can similarly
be observed in transit. A positive detection would give a further
major improvement in our understanding both of this specific system
and of planetary properties in general.

Rivera et al. (2005) carried out a photometric search for transits
by planet d ($P\sim1.974 {\rm d}$), and showed that such transits do
not occur. To date, photometric searches for transits by planet c
($P\sim30 {\rm d}$) and b ($P\sim60 {\rm d}$) have not been
reported. Although the \emph{a priori} geometric transit
probabilities for Gl 876 c and b are only $P_{\rm tr}=1.4\% $ and
$P_{\rm tr}=0.9\% $, respectively, the expected photometric transit
depths are greater than 10\%. Librational motions arising from the
resonance between the two outer planets lead to predicted transit
epochs for the outer planets that are not spaced evenly in time.
(The resonant structure of the system is discussed further in
Laughlin \& Chambers 2001; Rivera \& Lissauer 2001; Kley 2003; and
Beaug\'{e}, Michtchenko, \& Ferraz-Mello 2005).

Benedict et al. (2002) used HST astrometry to derive an orbital
inclination for Gl 876b of $i=84\pm 6^\circ $. Their result thus
suggests that planet b has a mass equal or close to the
RV-determined value of $M\sin (i)=1.89M_{\mbox{JUP}}$, and that the
system is within 1-sigma of being viewed edge-on. The implication of
the Benedict et al. (2002) measurement is that occurence of transits
is much more likely than the a-priori geometric probability would
suggest. The Benedict et al. (2002) result is in conflict, however,
with the dynamical models of the system presented by Rivera et al.
(2005), who find a best fit to the radial velocity data when the
system (assumed co-planar) is inclined by $i\sim50\pm3^{\circ}$ with
respect to the plane of the sky. Therefore, if transits are
detected, then the Benedict et al. (2002) results will be confirmed
at the expense of the conclusions of Rivera et al. (2005).

Gl 876b and Gl 876c are members of an as-yet unobserved class of
giant planets with intermediate surface temperatures. The expected
effective temperatures of ${T_{\rm eff}}_b =160$k and ${T_{\rm
eff}}_c =200$K of the planets are much lower than the value $T_{\rm
eff} \sim 1350$K measured with Spitzer for HD 209458b (Charbonneau
et al. 2005), but still considerably higher than the $T_{\rm eff}
\sim 135$K measured for Jupiter. Determination of their physical
properties would thus provide a useful link between Jupiter and
Saturn on the one hand, and objects such as HD 209458b on the other.
Accurate measurement of their radii would also give clues to their
interior structure (see Charbonneau et al. 2006).

Finally, the Gl 876 system with its strongly gravitationally
interacting planets and large planet-to-star radius ratios, has the
potential to display an extraordinarily informative set of transit
light curves. Our model light curves herein show what one can expect
when interacting transiting systems are discovered, for example,
with ongoing radial velocity surveys or with space-based missions
such as Kepler (Borucki et al. 2003) or COROT (Baglin 2003), and
allow for a more complete understanding of the multiplanet dynamics.

The plan of this paper is as follows. In \S 2, we describe our
dynamical model of the system, and illustrate some of the
photometric properties that the in-transit light-curves would be
expected to display. In \S 3, we describe an evaluation of the
Doppler radial velocity measurements that have been taken during
predicted transit intervals. In \S 4, we evaluate a range of
archival and newly obtained photometric data sets for evidence of
transits. In \S 5, we discuss our results, as well as the current
status and outlook for the ongoing {\it TransitSearch}
collaboration, and in \S 6, conclude.

\section{The Dynamical Model}
Gl 876 (M4V) is the fortieth-nearest star to Earth, with a
Hipparcos-determined distance of 4.69 pc (Perryman et al. 1997);
Tycho-II and UCAC positions validate this determination. As
discussed in Laughlin et al. (2005), current estimates of its mass
and radius stand at $M_\star =0.32\,M_\odot $, and $R_\star
=0.3\,R_\odot $. We adopt these values as fixed in this paper. To
date, 155 Doppler velocities measured at the Keck telescope and 16
velocities measured at the Lick telescope have been published for
the Gl 876 system (Marcy et al. 2001; Rivera et al. 2005). The two
planets induce a total velocity half-amplitude of the star of nearly
$0.25\,\mbox{km}\,\mbox{s}^{-1}$. Furthermore, as discussed in
Rivera et al. (2005) the system displays a very low level of stellar
radial velocity ``jitter'', which further aids in obtaining a
detailed characterization of the orbits. The system has now been
observed for more than eighty orbital periods of the middle planet,
c. This extensive time baseline allows for a much more detailed
study of planet-planet interactions than can be obtained with the
other exoplanetary systems known to be in 2:1 resonance, a list
which now includes HD 82943 (Mayor et al. 2004, Lee et al. 2006), HD
128311 (Vogt et al. 2005), and HD 73526 (Tinney et al. 2006). These
three systems all have inner planet periods of order ten times
longer than Gl 876 c. Our baseline co-planar $i=90^{\circ}$ orbital
model was obtained by Rivera et al. (2005) from a dynamical fit to
the Keck radial velocity data. For reference, the parameters of this
model are reproduced in Table 1.\footnote{Note that the elements
reported in Table 1 are osculating orbital elements (see, e.g.,
Murray \& Dermott 1999) referenced to the epoch JD 2452490.}
Additional fits with $i<90^{\circ}$ were also obtained to the Keck
data using the algorithm described by Rivera et al. (2005). As with
the $i=90^{\circ}$ baseline model, the system is assumed to contain
three planets, all orbiting the star in the same plane. We varied
the inclination of the orbital plane to the plane of the sky from
$i=90^{\circ}$ to $i=89.0^{\circ}$ in decrements of $0.1^{\circ}$,
and fitted for each of the 13+1 remaining parameters. N-body
integrations of the fit then generate a sequence of transit
predictions for the outer two planets. Uncertainties in the
predicted transit midpoints were generated with the bootstrap
Monte-Carlo technique described by Rivera et al. (2005). For both
outer planets, the uncertainty in the transit midpoint times was
found to vary from $\sim$1 hr to a few hours. The variation in
uncertainty from transit to transit is plotted in Figure 1. The
predicted transit midpoints are best constrained during the epoch
spanned by the radial velocity observations. Errors increase as the
transit predictions are generated for earlier or later times. Light
curves were produced with a photometric model of the transit which
assumes that the planets are opaque disks with radii $R=0.93 \,
R_{\rm JUP}$ for c and $R=1.04 \, R_{\rm JUP}$ for b. These
equatorial radii are computed using the models of Bodenheimer,
Laughlin, \& Lin (2003). The specific intensity of the stellar disk
is modeled with a linear limb darkening law, given as

\begin{equation}
I(\mu)/I(1)=1-v(1-\mu) , \label{eq1}
\end{equation}

where $I$ is the specific intensity of the stellar disk, $v=0.724$
is the corresponding limb darkening coefficient appropriate to
V-band observations of an M4V primary (per Claret 2000); and $\mu =
\cos(\theta)$.

Figures 2 and 3 show a selection of model light curves for planets b
and c. For planet b, which has an eccentric ($e=0.22$), rapidly
precessing orbit ($d\varpi/dt\sim44^{\circ}\,{\rm yr}^{-1}$), the
shape of the predicted curves depends strongly on the osculating
periastron angle. For $i<90^{\circ}$, the transits are both shorter
and deeper when the transits occur near periastron. The orbital
eccentricity of planet b is more nearly circular, with $e=0.025$. In
this case, there is little variation in the predicted light curves
as the orbits precess. The known transiting planets have
transit-to-transit intervals $\Delta T_{i}$ that are evenly spaced
to within the current resolution of the observations. HD 209458b,
for example, has had individual transit midpoints measured to an
accuracy of several seconds (Brown et al. 2001, Wittenmyer et al.
2005), and an average period measured to an accuracy of 83 ms
(Wittenmyer et al. 2005, Winn \& Holman 2005). If, however, one is
able to measure differences in the time intervals between successive
transits, one can derive additional dynamical information. As
discussed by Miralda-Escud\'{e} (2002), Holman \& Murray (2005), and
Agol et al. (2005), measurements of small variations from exact
periodicity can be used to infer the presence of additional bodies
in the system, potentially with the masses in the range appropriate
to terrestrial planets. In the event that either Gl 876b or Gl 876c
were to transit, one would expect the departures from strict
periodicity to be large. In fact, successive transits of planet c
would occur at intervals differing by as much as 4 hours! By fitting
a dynamical model to such transit intervals, one could obtain
information regarding all of the orbital elements of the planets
(including inclinations and nodes).

\section{Seeking Evidence of Transits in the Radial Velocity Data Set}
During a planetary transit, the occulting disk of the planet passes
over portions of the stellar disk containing a varying component of
line-of-sight velocity arising from stellar rotation; this phenomena
produces an asymmetry in the star's spectral lines known as the
Rossiter-McLaughlin effect (Rossiter 1924, McLaughlin 1924). The
effect has been described in detail in the context of planetary
transits by Ohta, Taruya \& Suto (2005), and has been analyzed for
HD 209458 by Queloz et al. (2000), Bundy \& Marcy (2000), and Winn
et al. (2005), and for HD 149026 by Wolf et al. (2006). In fact, the
transiting planet HD 189733 was initially discovered to transit
\textit{not} via photometry, but rather via such in-transit radial
velocity variations, found by Bouchy et al. (2005). For our study of
Gl 876, we compared the observation epochs of all of the Lick (see
Marcy et al. 2001) and Keck (see Rivera et al. 2005) radial velocity
observations to see if any of these observations occurred during a
predicted transit (assuming $i=90^{\circ}$). For Gl 876, we used a
rotational period that was determined to be 96.7 days using
rotational modulation (as reported in Rivera et al. 2005); we used
this value to calculate the amplitude of the Rossitter-McLaughlin
effect. Four such observations were found, three taken during a
predicted transit by planet ``c'', and one taken during a predicted
transit by planet ``b''. Figures 4 and 5 display the relevant parts
of the radial velocity model (both with and without the
Rossiter-McLaughlin effect), with the observations shown. The radial
velocity model curves in both figures were generated by adding
template curves which use the stellar and planetary radii and limb
darkening law given above (and which apply the analytic integrals
given in Ohta et al. 2005), to our fiducial $i=90^{\circ}$
three-planet radial velocity model. For the transit epoch in the
case of planet c (Figure 4), the data is inconsistent with a
transit. In Figure 5, for the outer planet b, the situation is
unresolved.

\section{A Search for Transits in Photometric Data Sets}

We have also compared our model light curves to a variety of
photometric datasets for Gl 876 taken since 1989. The results of
these comparisons are as follows.

{\bf Hipparcos}: The Hipparcos Mission (see Perryman et al. 1997)
produced 67 accepted photometric measurements for Gl 876 (HIP
113020). These measurements have a median magnitude $H_{\rm
P}=10.148$, with $\sigma=0.0045$. There is evidence for a weak
periodicity when the data is folded at the 96.7 {\rm d} rotational
period of the star (also noted in Rivera et al. 2005). In Figure 6,
we plot our model $i=89.6^{\circ}$ light curves for Gl 876 b and c
superimposed on the Hipparcos epoch photometry. None of the
Hipparcos measurements were made near or during a predicted transit
window.

{\bf Fairborn Observatory}:

A number of long-term photometric monitoring programs are being
carried out with the automated telescopes at Fairborn Observatory
(e.g., Henry 1999;  Eaton, Henry, \& Fekel 2003) and, as described
in detail in Rivera et al. (2005), Gl 876 has been on one of the
observing programs. Figures 7 and 8 show the Fairborn Observatory
data and our corresponding model light curve for planets b and c.
Data taken during predicted transit epochs for planet c show no
indication of a transit. The Fairborn Observatory data indicate that
long-term stellar variability from Gl 876 is of order 0.05
magnitude, which is somewhat smaller than the variation observed
with Hipparcos.

{\bf Transitsearch and AAVSO}: Seagroves et al. (2003) describe {\it
TransitSearch}, a cooperative distributed observing project
involving sub-meter class telescopes worldwide. The TransitSearch
strategy is to observe known planet-bearing stars at the dates and
times when transits are expected to occur. It is therefore a
targeted search, which differentiates it from ongoing wide-field
surveys. We identify observing windows for candidate stars through
the use of the bootstrap Monte Carlo technique described by Laughlin
et al. (2005). The observational campaigns are prioritized by the
a-priori likelihood that a particular candidate planet will display
transits. This likelihood is given by,

\begin{equation}
\textsl{P}_{transit} =0.0045\left( {\frac{1AU}{a}} \right)\left(
{\frac{R_\ast +R_{pl} }{R_\odot }} \right) \left[\frac{1+e\cos({\pi
\over{2}}-\varpi)}{1-e^{2}}\right]\, \label{eq2}
\end{equation}

where \textit{a} is the semi-major axis of the orbit, $R_\ast$ is
the radius of the star, $R_{pl}$ is the radius of the planet,
$\varpi$ is the longitude of periastron, and $e$ is the orbital
eccentricity.

The first TransitSearch dataset for Gl 876 was obtained by one of us
(Shankland) from 2003 Oct. 27.9583 to 2003 Oct. 28.0368, with a
V-filtered, 16-bit speckle ccd at the primary focus of a 40.4 cm
f/5.1 reflector (at air mass from $\sec(z)=2.7$ to 1.7). The
photometric time series shows a monotonic flux increase of 0.16 +/-
0.04 mag from 2003 Oct. 27.9583 to 2003, Oct 27.9972. As shown on
the left panel of Figure 9, the depth, duration and timing of this
event were consistent with an egress from transit of Gl 876 c, in
agreement with our predictions. The possibility that a transit had
been observed spurred us to orchestrate a distributed observational
campaign during the 2004 season.

During the 2004 June 24 opportunity, observations were attempted by
ten observers in Australia. Poor weather thwarted all but one
observer, and no transit signal was seen in the data, which covered
about half of the $1-\sigma$ transit window. Eight Australian and
Japanese observers participated in the following 2004 July 24 planet
c campaign. They were all clouded out. Weather improved for the 2004
August 23 planet c opportunity, permitting observations by five of
the nine Australian, South African, and German observers. No transit
signal was seen with photometric depth greater than 1\%.

In 2004 October, a number of AAVSO and TransitSearch observers
obtained photometry during a closely spaced pair of transit windows
for planets b and  c. Beginning October 20 11:22 UT and ending
October 23 15:27 UT, a total of 2,795 CCD observations were
obtained. In addition, 2,981 photometric observations were obtained
in the days immediately before and after the window to set baseline
activity and to look for red dwarf flaring. Such a large,
distributed network of observers allowed coverage of the window with
very few observing gaps over the 3-day period. Gaps in the coverage
were smaller than the estimated transit period or forecast window,
allowing us to conclude that transits for planets b and c did not
occur during the October 2004 campaign. Photometric data from the
campaign are plotted in Figures 9 and 10.

Individual observations and uncertainty estimates from the October
2004 campaign are available from the AAVSO web site, while the other
datasets can be found at the {\it TransitSearch}
website\footnote{our AAVSO data available at
http://www.aavso.org/data/download/ while other collected data can
be found at
http://www.ucolick.org/\%7elaugh/Gl876/\_\_\_\_c.results.html}. We
are continuing to obtain photometry of Gl 876, with the goal of
gaining a better understanding of the photometric variability of
mature (age exceeding 1Gyr) red dwarfs.

\section{Discussion}

The {\it TransitSearch} photometric network has been in operation
for three years, and has participating observers capable of
providing fully global longitude and latitude coverage. The network
maintains a continuously updated catalog\footnote{available at
http://www.ucolick.org/\%7elaugh/} of the known census of extrasolar
planets. The catalog is presently the only available source for (1)
a-priori transit probabilities and predicted transit depths, for the
known extrasolar planets, (2) predicted transit ephemeris and
transit window uncertainties based on orbital fits to published
radial velocities, and (3) the results of known photometric searches
for transits of planet-bearing stars. To date, the majority of
negative transit searches have not been reported in the literature.
With this paper, we hope to spur a reversal of that trend.

Due to significant uncertainties in the orbital fits, it is
generally very difficult to definitively rule out transits by
planets with $P>10 {\rm d}$. The Gl 876 planets form a notable
exception to this rule because of their high signal-to-noise,
$\Gamma\sim50$, and the extensive radial velocity data set allows
for a very accurate orbital fit. Campaigns by the Transitsearch
collaboration have thus far led to transits of a number of other
shorter-period planet-bearing stars being ruled out to a high degree
of confidence. The planets for which the possibility of transits
have been significantly discounted by our campaigns include HD
217107 ``b'' ($P=7.127 {\rm d}$, $P_{\rm tr}=8.0\%$, see also Vogt
et al. 2005), HD 168746 ``b'' ($P=6.403 {\rm d}$, $P_{\rm
tr}=8.1\%$, see also Pepe et al. 2002), and HD 68988 ``b'' ($P=6.276
{\rm d}$, $P_{\rm tr}=9.0\%$, see also Vogt et al. 2002).

Furthermore, planets for which usable photometry of the parent stars
has been obtained by our network during the $3-\sigma$ transit
windows, but for which transits can not yet be fully ruled out,
include HD 188753Ab ($P=3.348 {\rm d}$, $P_{\rm tr}=11.8\%$), HD
76700 ``b'' ($P=3.971 {\rm d}$, $P_{\rm tr}=10.0\%$), HD 13445 b
($P=15.76 {\rm d}$, $P_{\rm tr}=3.4\%$), HD 74156 ``b'' ($P=51.64
{\rm d}$, $P_{\rm tr}=3.8\%$), HD 37605 ``b'' ($P=54.23 {\rm d}$,
$P_{\rm tr}=2.2\%$), and HD 80606 ``b'' ($P=111.4 {\rm d}$, $P_{\rm
tr}=1.7\%$). Planets for which campaigns are scheduled to begin soon
include GL 581 ``b'' ($P=5.366 {\rm d}$, $P_{\rm tr}=3.6\%$), and HD
99492 ``b'' ($P=17.04 {\rm d}$, $P_{\rm tr}=3.4\%$). GL 581 is of
particular interest because the primary star is an M dwarf, and the
planet (if it transits) will be a Neptune-mass object
$M\sin(i)=0.052 M_{\rm Jup}$ (Bonfils et al. 2005). The 1.6\%
transit depth would allow an unprecedented physical characterization
of a low-mass extrasolar planet. To date, our Gl 876 campaign has
employed the most globally far-reaching distribution, and along with
recent leaps in fidelity in similar campaigns, clearly underscores
the utility of the network.

What is more, Gl 876 b and c constitute a dramatic exception to the
emerging aphorism that Jupiter-mass planets are rarely associated
with M dwarfs. Indeed, a paucity of giant planets orbiting red
dwarfs seems to be a natural consequence of the core accretion
theory of planet formation (Laughlin, Bodenheimer \& Adams 2004).
The core-accretion theory predicts, however, that Neptune-mass and
smaller mass planets will be very common around red dwarfs. This
stands in contrast to the predictions of Boss (2006), who suggests
that the gravitational instability mechanism is the dominant mode of
giant planet formation. If this is true, then Jupiter-mass planets
should be just as common around red dwarfs as they are around
solar-type stars. Efforts to detect transits of small planets
orbiting small stars are therefore likely to gain increasing
importance and focus over the next several years.

There are, in fact, several observational indications that low-mass
planets may be common around red dwarfs. The radial velocity surveys
have recently reported the detection of Neptune-mass companions in
short-period orbits around the red dwarfs GL 581 and GL 436. The
OGLE team of observers using the microlensing method have detected
the signature of what appears to be a planet with 5.5 Earth Masses
orbiting a distant red dwarf (Bennet et al. 2006). With these points
in mind, we can put the results of our Gl 876 campaign in a broader
context. A ground-based (and ``fiscally viable'') photometric
network such as Transitsearch, especially in collaboration with a
dedicated bank of telescopes, can capably monitor individual M
dwarfs to search for transits of terrestrial-sized planets. Owing to
the intrinsic long-term variability of M dwarfs, candidate planets
would have to be identified on the basis of a full transit signature
in a single time-series.

A 0.1 $M_{\odot}$ M dwarf has $R_{\star}/R_{\odot}\sim0.1$, $T_{\rm
eff}\sim2750 {\rm K}$, and $L_{\star}/L_{\odot}\sim5\times10^{-5}$;
a habitable planet that receives an Earth-equivalent flux from the
star thus needs to orbit at a distance of just 0.022 AU, which
corresponds to an orbital period of 3.85 days. An Earth-like planet
with such a period would be rotationally synchronized to the red
dwarf, and in the absence of any significant perturbing bodies, its
orbit would be almost perfectly circular. Interestingly however,
simulations show that habitability is unlikely to be adversely
affected by a spin-orbit period synchronization. Joshi, Haberle, \&
Reynolds (1997) used a global climate model to investigate how the
Earth's climate would respond if the Earth were tidally locked to
the Sun, and found that Earth remains habitable in this
configuration -- at 1 AU, and perhaps throughout the habitable zone.

The $M=7.5 M_{\oplus}$, $P=1.9379 \,{\rm d}$ companion to Gl 876
demonstrates that it is not unreasonable to expect terrestrial-mass
bodies on short-period orbits of $0.1 M_{\odot}$ stars. Montgomery
\& Laughlin (2006) have carried out accretion simulations with the
Wetherill-Chambers method (Wetherill 1996, Chambers 2001) that model
the accretion of terrestrial-mass planets in short-period orbits
about M dwarfs. These simulations provide further evidence that
terrestrial-mass planets will commonly form in the desired
short-period orbits.

If habitable planets do commonly form in orbit around low mass M
dwarfs, the chances of detecting and characterizing them are
surprisingly good. The transit of an Earth-sized planet will block
about 1\% of the stellar flux of a $0.1 M_{\oplus}$ star. For a
planet on a habitable 3.85 day orbit, the transit will be relatively
brief, $\sim40$ minutes. The a-priori geometric probability of
observing such a transit is 2\%.

A 1\% photometric dip is readily detectable. Amateur astronomers who
participate in the Transitsearch collaboration routinely achieve
detection thresholds of considerably better than 1\%, as evidenced
by confirming detections of HD 149026b which has a transit depth of
just 0.3\%. Indeed, capable amateur observers have demonstrated the
photometric capability to detect the passage of a Mars-sized body in
front of an 11th magnitude 0.1 solar mass red dwarf.

Inevitably one must ask how many suitable red dwarfs are available
on the sky as a whole. Even though the lowest-mass M stars are the
most common type of star, they are also exceedingly dim. A dedicated
3m-class telescope would be required to properly search the
magnitude range V=16-18, where observations of the lowest-mass red
dwarfs begin to be plentiful (and indeed, many 0.1 $M_{\odot}$ stars
within 10 parsecs still likely remain to be discovered). In the
meantime, however, we recommend that observers obtain high-cadence
photometry of nearby single stars listed in the RECONS
catalog\footnote{maintained by T. Henry and collaborators at Georgia
State University.} of the 100 nearest stellar systems; these include
Proxima Centauri, Barnard's Star, Wolf 359, Ross 154, Ross 128, DX
Cancri, GJ 1061, GJ 54.1, and GJ 83.1. We estimate that each one has
a $\sim1$\% chance of harboring a detectible transiting, potentially
habitable planet - readily detectible by distributed photometric
observation.

\section{Conclusions}

Our study of Gl 876 has produced several interesting results. We
have demonstrated that transits of planets b and c are not currently
occurring, and we have thus modestly constrained the parameter space
available to observationally allowed orbital configurations of the
Gl 876 system. The absence of transits is consistent with the
dynamical study of Rivera et al. (2005) who find that a co-planar
system inclination $i\sim50^{\circ}$ yields the best fit to the
radial velocity data. Our dynamical analysis indicates that when
planetary transits in a strongly interacting system similar to Gl
876 are eventually observed, then a tremendous amount of information
will be obtained. We look forward to the discovery of the first such
system.

We have also shown that a globally distributed network of
small-telescope observers can effectively provide definitive
photometric follow-up of known planet-bearing low-mass stars. Our
experience with Gl 876 indicates that intensive photometric
monitoring campaigns of individual, nearby M dwarfs constitute a
tractable strategy for planet detection. The Sun's nearest low-mass
stellar neighbors should therefore be scrutinized very carefully for
transits by short-period terrestrial planets. A low-budget survey of
this type provides a small, but nevertheless non-negligible chance
to discover a truly habitable world from the ground, and within a
year or two. Such a detection would be a remarkable and exciting
discovery.

\subsection{Acknowledgements}
We thank K. Johnston, R. Gaume, K. Malatesta, J. Lazio, D. Boboltz,
C. Shaw, B. Casey, E. Albert, A. Henden, J. Larsen, D. Katz, C.
Phillips, W. Bollwerk, Z. Dugan, J. Jaglowicz, and T. Castellano for
useful discussions or other support.

We are grateful to the AAVSO and its many members who provided
campaign support in addition to those authors listed above. We also
gratefully acknowledge the insightful suggestions and constructive
feedback offered by our referee.

This material is based upon work supported by the National
Aeronautics and Space Administration under Grant NNG04GN30G issued
through the Origins of Solar Systems Program to GL. GWH acknowledges
support from NASA grant NCC5-511 and NSF grant HRD-9550561. This
research has made use of the Simbad, USNO and {\it TransitSearch}
databases, operated at CDS, Strasbourg, France, USNO, Washington,
D.C., USA, and UCSC, Santa Cruz, California, USA, respectively.

\clearpage

\begin{figure}[htp]
\centering
\includegraphics[angle=-90,width=0.99\textwidth,viewport=0 0 600 800]{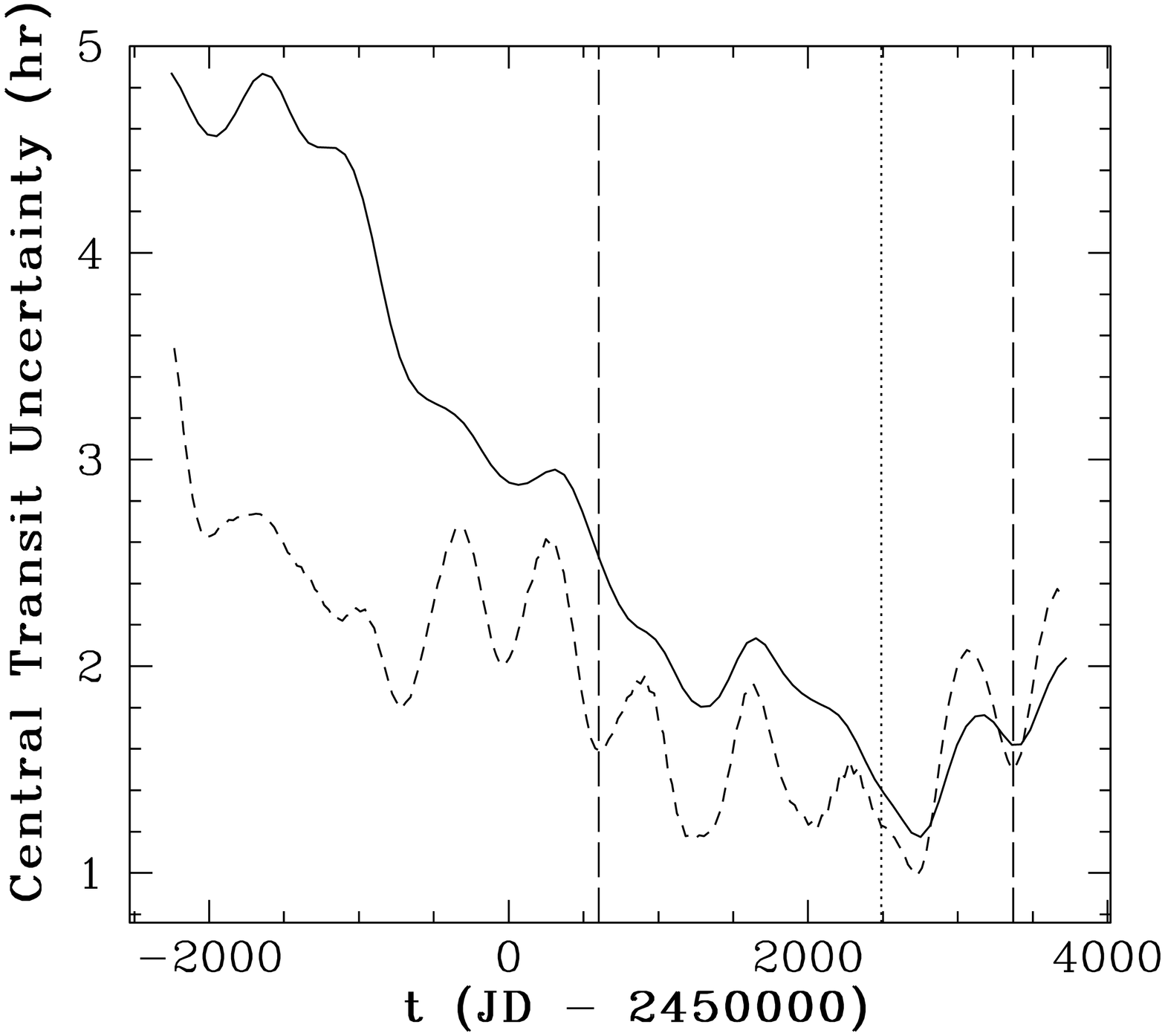}\\
\caption{ Uncertainty in the central transit time for planets ``b''
{\it solid line} and ``c'' {\it dashed line}. The vertical dashed
lines delineate the time span of the Keck RV observations used to
generate the co-planar fits that form the basis of our model
photometric and RV curves. The vertical dotted line indicates the
epoch for the fits (JD 2452490.0). }
\end{figure}

\begin{figure}
\includegraphics[angle=-90,width=0.99\textwidth,viewport=0 0 600 800]{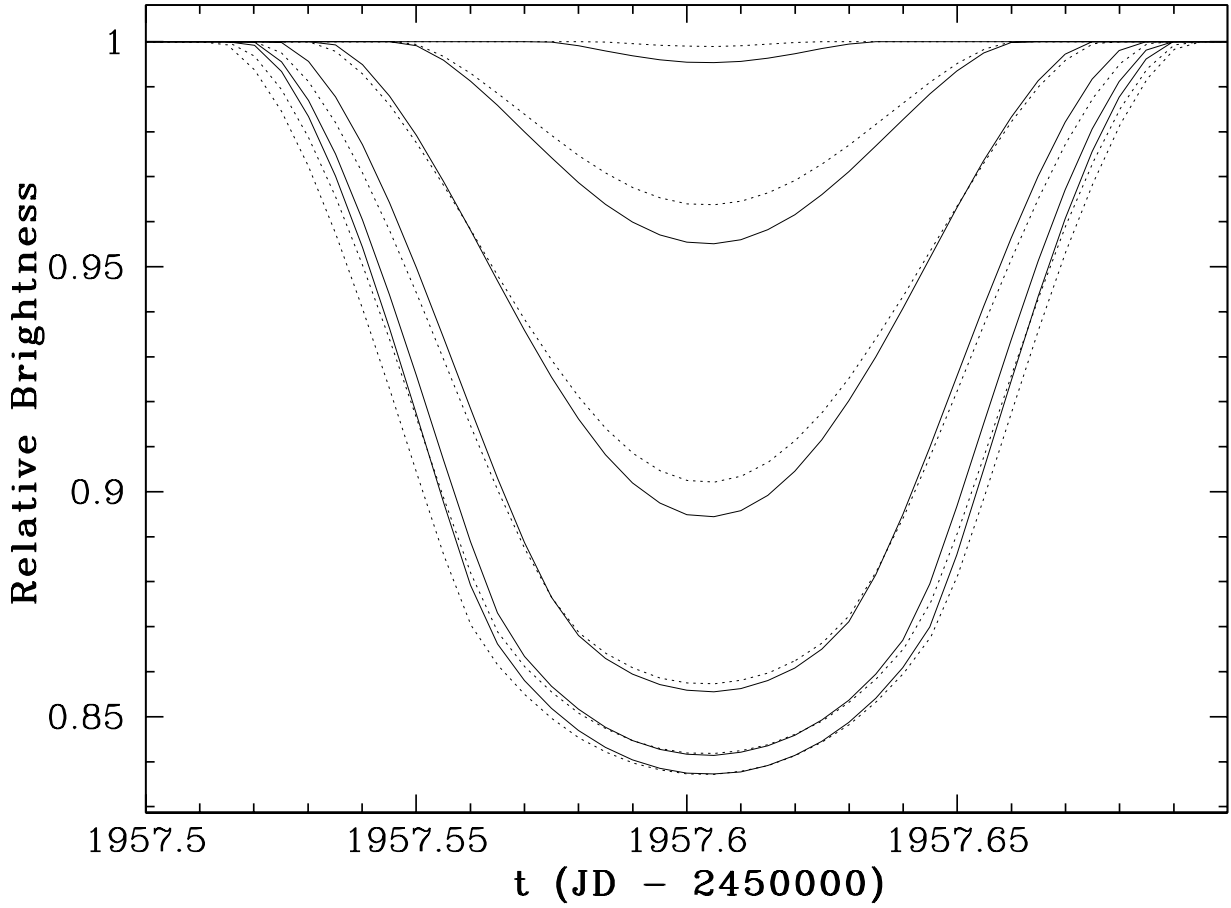}\\
\caption{ Model limb-darkened light curves during a
predicted transit by planet ``b'' when it is near periastron
(\emph{solid curves}) and near apastron (\emph{dashed curves}). The
curves correspond to different assumed inclinations ($i=90^{\circ}$,
$89.9^{\circ}$, $89.8^{\circ}$, ..., $89.5^{\circ}$). The shallowest
occur for $i=89.5^{\circ}$.}
\end{figure}

\begin{figure}
\includegraphics[angle=-90,width=0.99\textwidth,viewport=0 0 600 800]{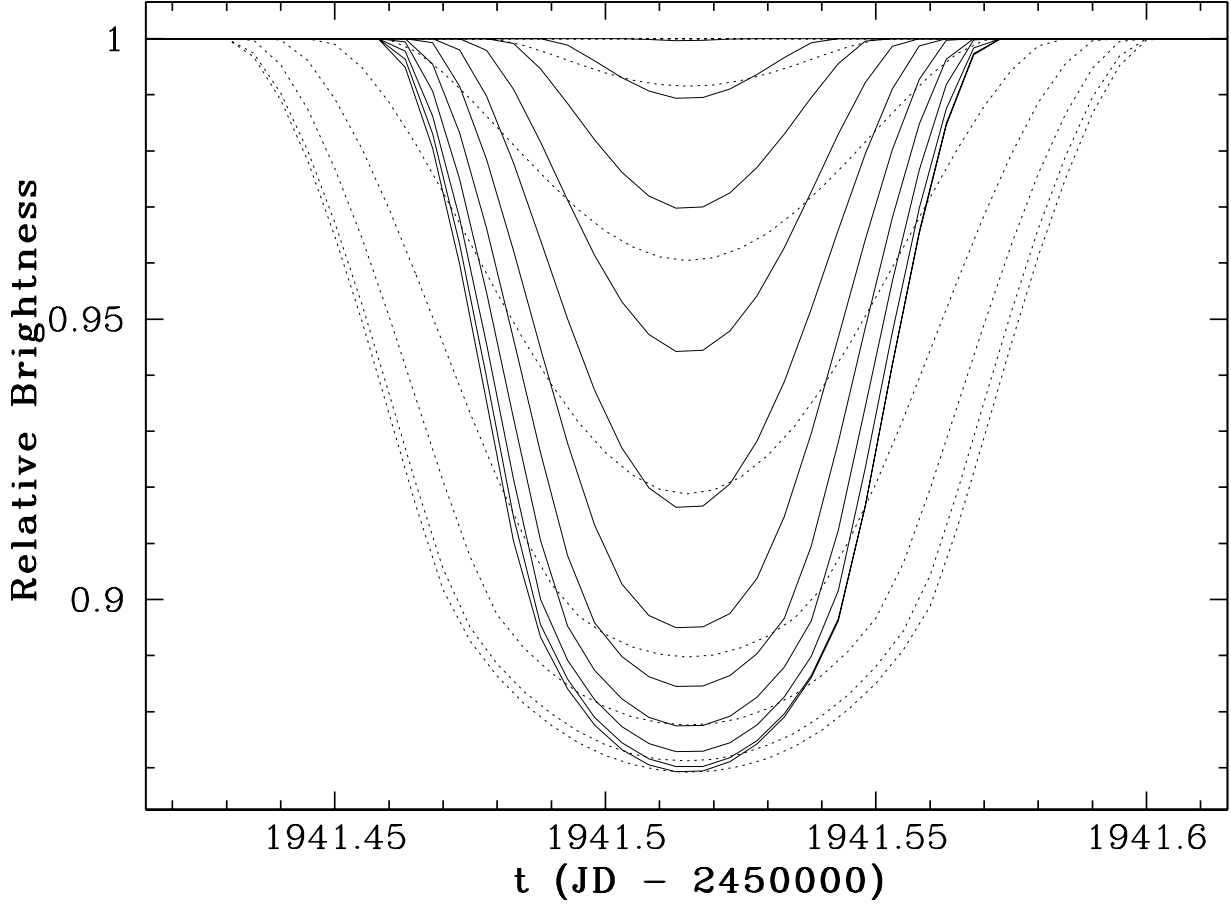}\\
\caption{ Model transit light curves for planet
``c'', when it is near periastron ({\it solid lines}), and near
apastron ({\it dashed lines}). Inclinations, $i=90^{\circ}$ through
$i=89.0^{\circ}$ are shown in decrements of $0.1^{\circ}$. Planet
``c'''s eccentricity $e\sim0.22$ leads to a greater difference in
transit depth and duration between periastron and apastron. }
\end{figure}

\begin{figure}
\includegraphics[angle=-90,width=0.99\textwidth,viewport=0 0 600 800]{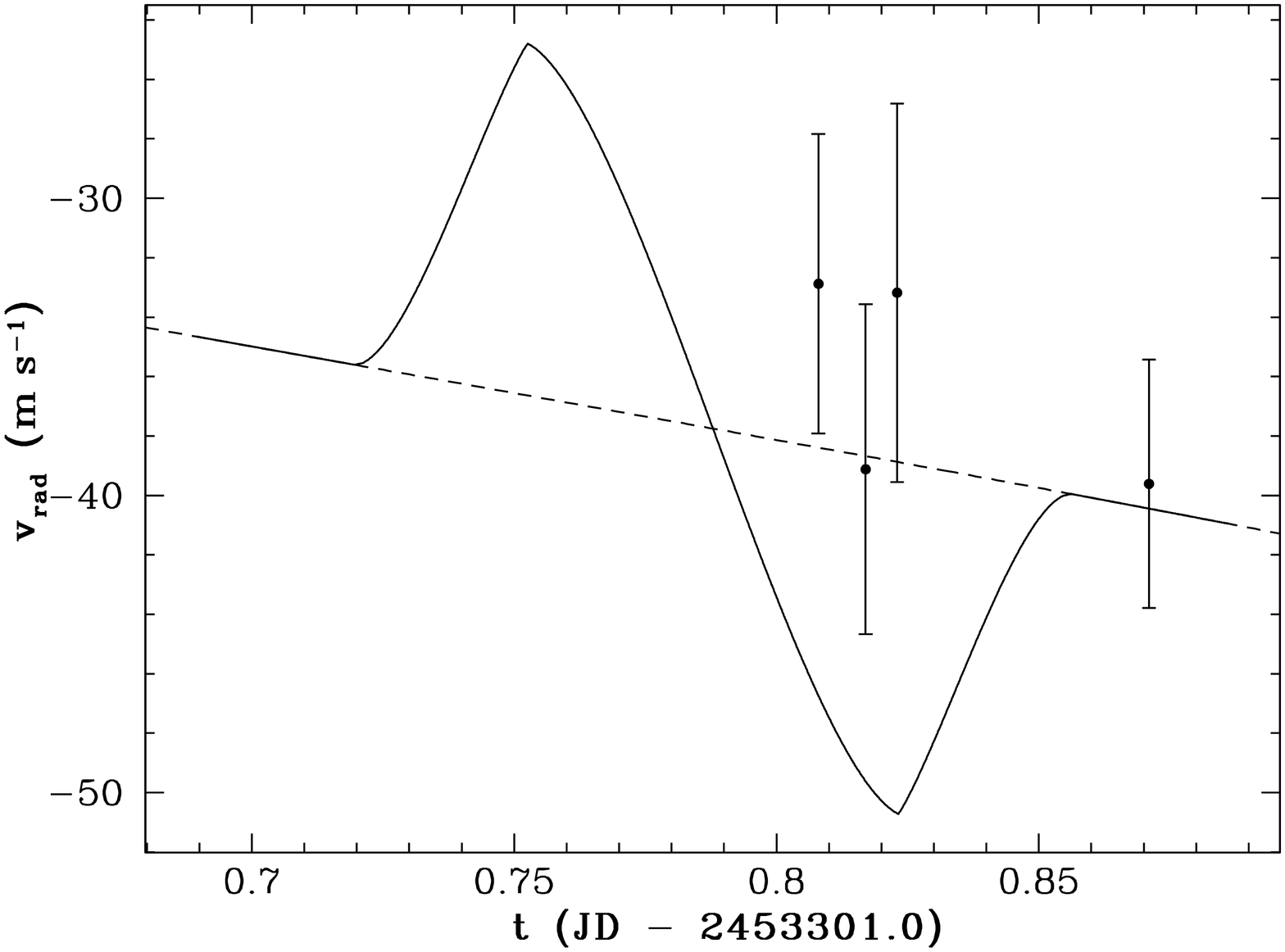}\\
\caption{ Three-planet RV model (\emph{dashed
curve}) with a superimposed model Rossiter curve (solid line) during
a predicted transit by planet ``c'' near epoch JD 2453301.79. The
observed Keck RV data is shown as the black filled-in circles with
error bars. }
\end{figure}

\begin{figure}
\includegraphics[angle=-90,width=0.99\textwidth,viewport=0 0 600 800]{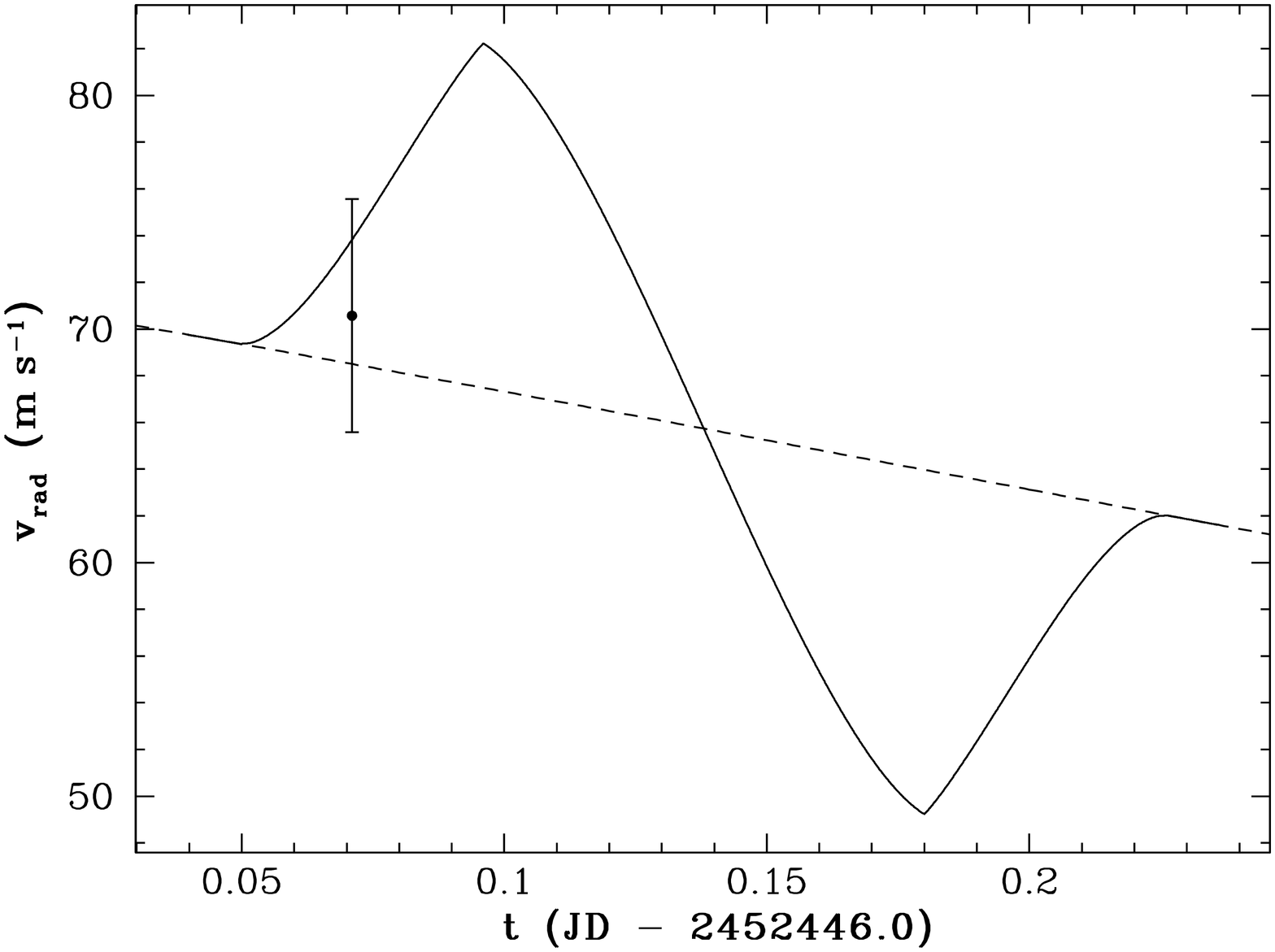}\\
\caption{ Three-planet RV model (\emph{dashed
curve}) with a superimposed model Rossiter curve (solid line) during
a predicted transit by planet``b'' near epoch JD 2452446.14. Again,
the observed Keck RV data is shown as the black filled-in circle
with error bars. }
\end{figure}

\begin{figure}
\includegraphics[angle=-90,width=0.99\textwidth,viewport=0 0 600 800]{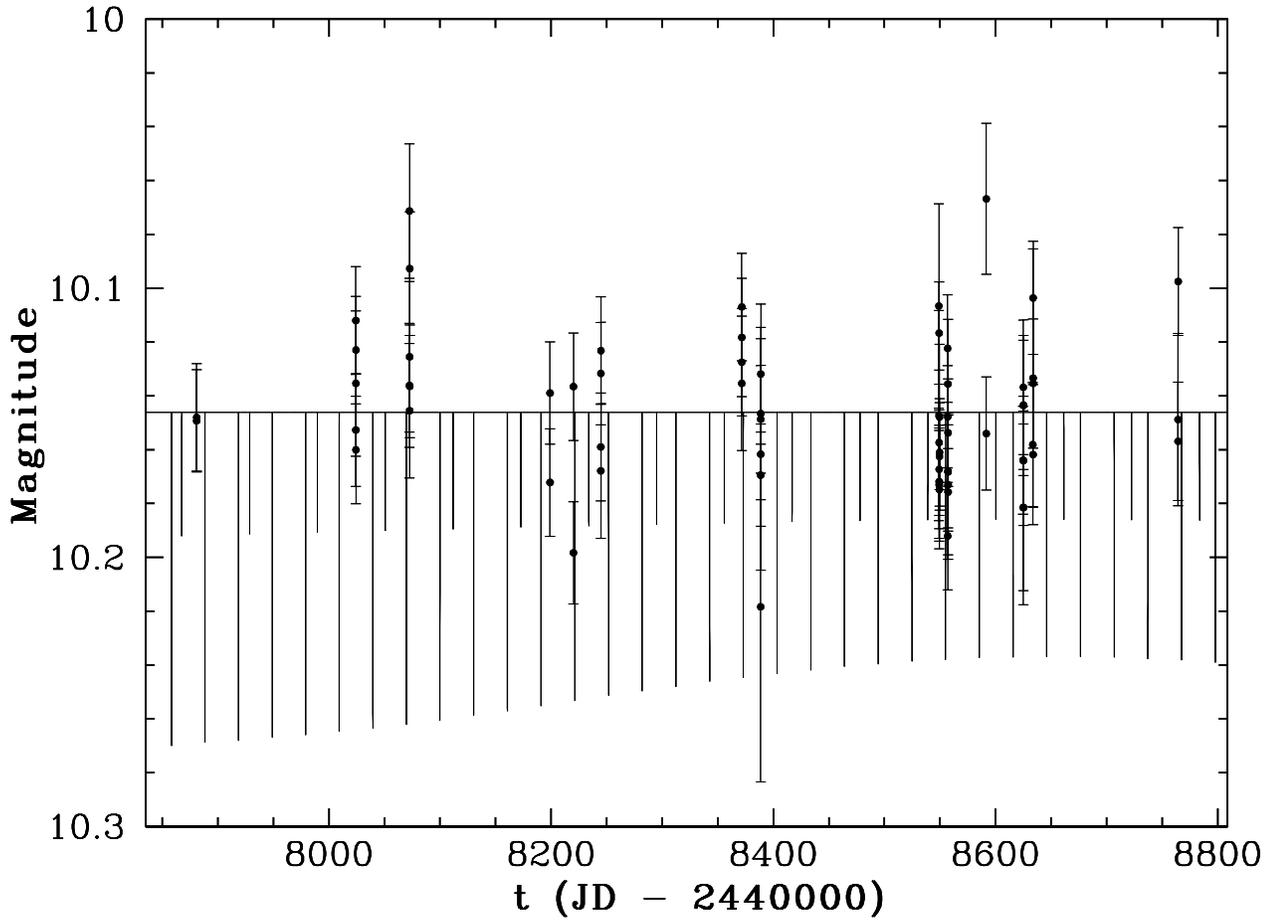}\\
\caption{ Model $i=89.6^{\circ}$ light curves for Gl
876 ``b'' and
 ``c''  superimposed on Hipparcos epoch photometry. None of the
Hipparcos measurements were made near or during a predicted transit
window. }
\end{figure}

\begin{figure}
\includegraphics[angle=-90,width=0.99\textwidth,viewport=0 0 600 800]{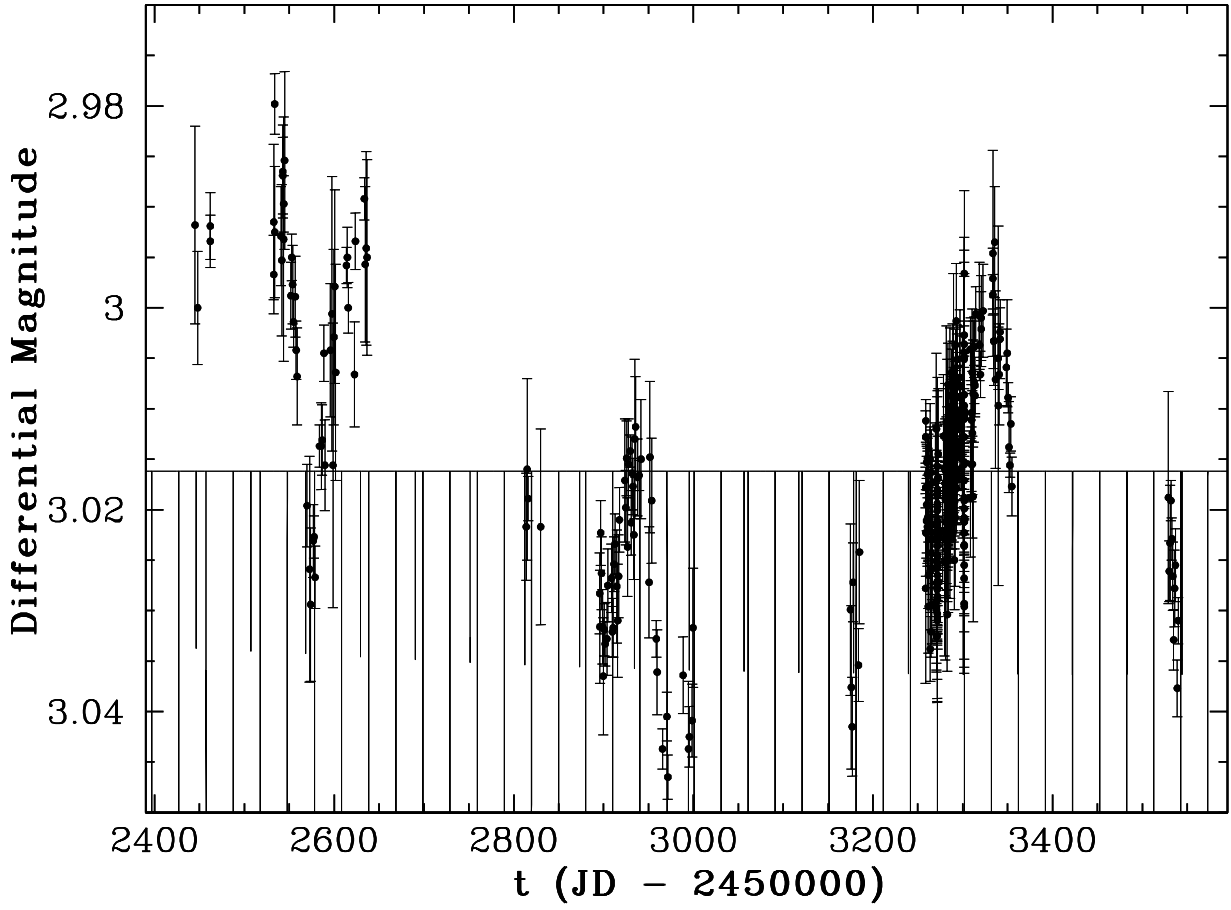}\\
\caption{ Model $i=89.3^{\circ}$ light curves for Gl
876 ``b'' and ``c'' superimposed on Fairborn Observatory photometry.
Note that the intrinsic long-term variability for Gl 876 over a
$\sim1000$ day period is of order 0.05 magnitudes. }
\end{figure}

\begin{figure}
\includegraphics[angle=-90,width=0.99\textwidth,viewport=0 0 600 800]{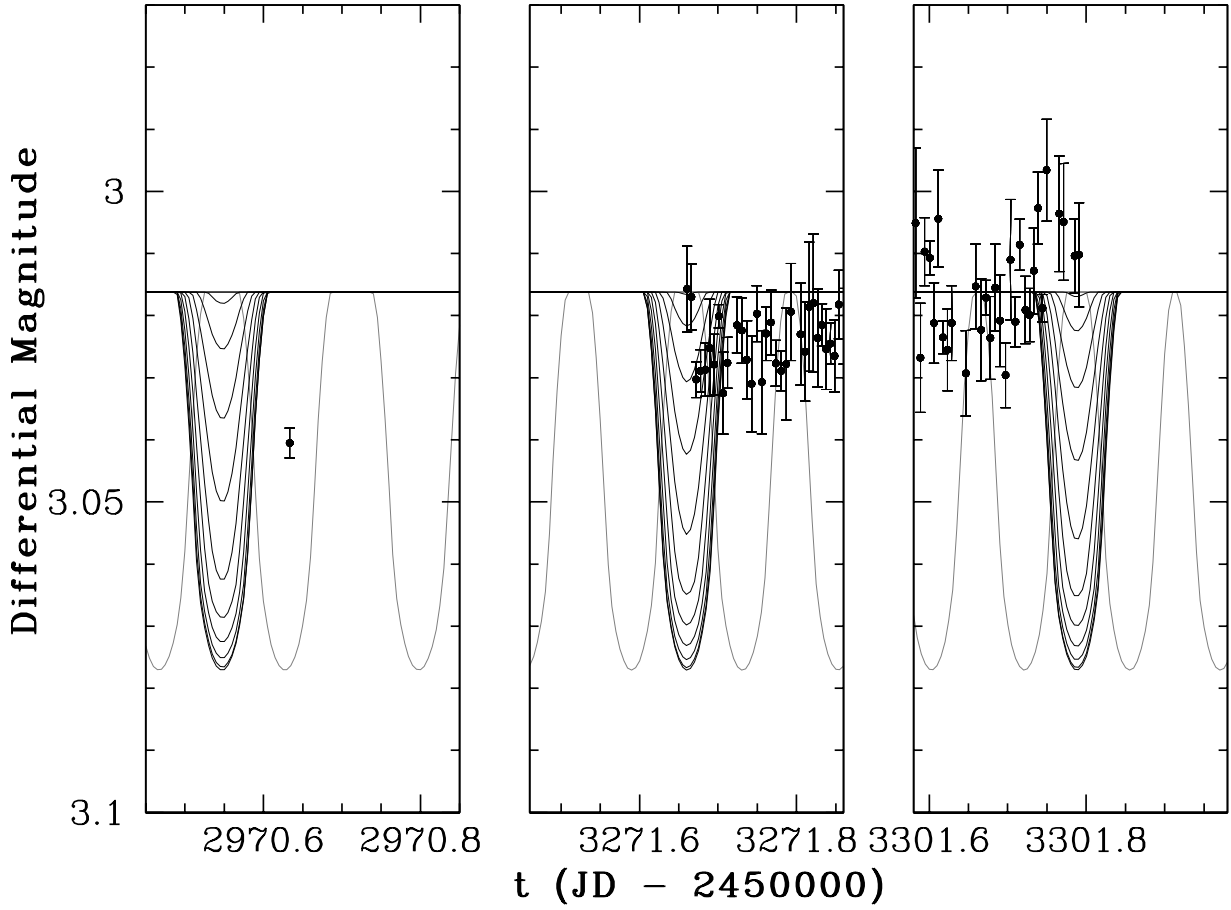}\\
\caption{ Model light curves for Gl 876 ``c''
superimposed on Fairborn Observatory photometry taken near three
individual predicted transit windows. The solid lines correspond to
model light curves arising from assumed inclinations running from
$i=90.0^{\circ}$ to $i=89.0^{\circ}$. The light lines correspond to
$i=90^{\circ}$ model light curves arising from bootstrap fits in
which the transit is $3\sigma$ early, $1\sigma$ early, and $1\sigma$
and $3\sigma$ late. }
\end{figure}

\begin{figure}
\includegraphics[angle=90,width=0.99\textwidth,viewport=0 0 600 800]{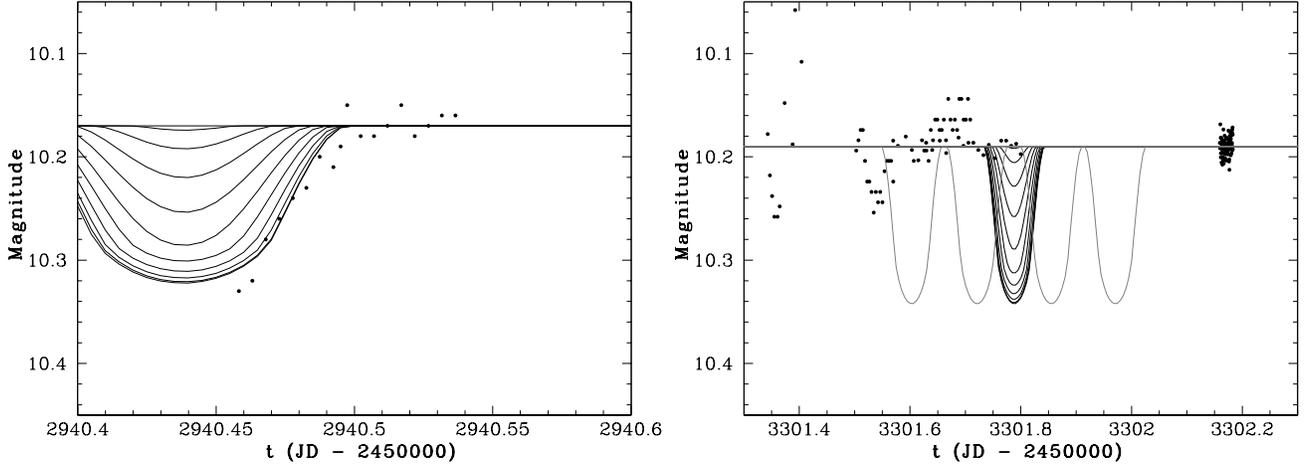}\\
\caption{ {\it Left: }Shankland 2003 data for ``c'',
this plate shows dynamical model light curves again for the same
decrements of $i$, taken near JD 2452940.4381; relative photometry,
which began in astronomical twilight (\emph{filled-in circles}) {\it
Right: } Dynamical model light curves for Gl 876 due to planet ``c''
for $i=90.0^{\circ}$ through $89.0^{\circ}$ in $0.1^{\circ}$
decrements, taken near a predicted transit centered at JD
2453301.7879. The data were compiled, baselined and shifted to
employ a mean magnitude of V=10.19, based on relative photometry
obtained by AAVSO and {\it TransitSearch} (\emph{filled-in
circles}). The light lines correspond to $i=90^{\circ}$ model light
curves arising from bootstrap fits in which the transit is 3$\sigma$
early, 1$\sigma$ early, and 1$\sigma$ late and 3$\sigma$ late.}
\end{figure}

\begin{figure}
\includegraphics[angle=-90,width=0.99\textwidth,viewport=0 0 600 800]{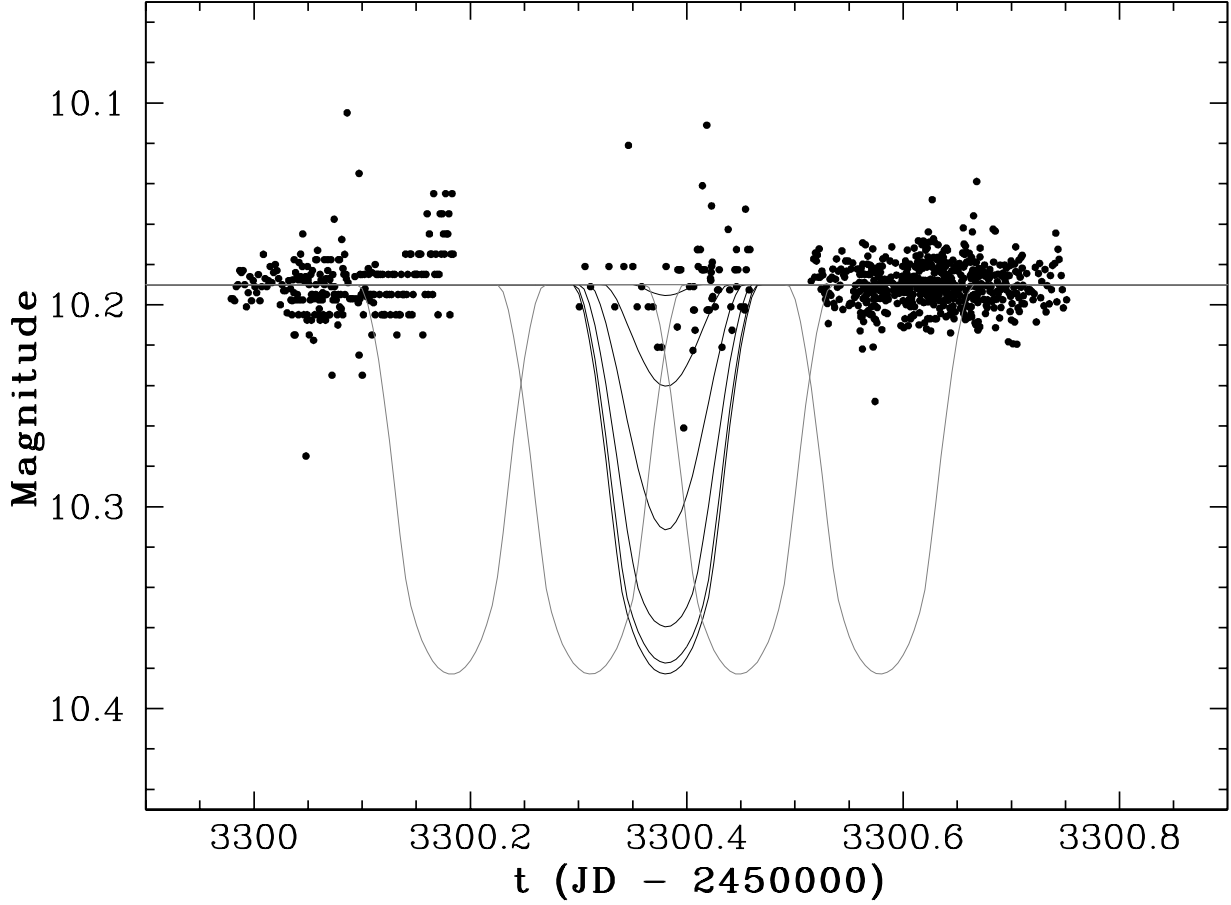}\\
\caption{ Dynamical model light curves for Gl 876
due to planet ``b'' for $i=90.0^{\circ}$ through $89.5^{\circ}$ in
$0.1^{\circ}$ decrements, near a predicted transit centered at JD
2453300.3806. As with the previous figure, the data were baselined
and shifted to employ a mean magnitude of V=10.19, based on relative
photometry obtained by AAVSO and {\it TransitSearch}(\emph{filled-in
circles}). The light lines correspond to $i=90^{\circ}$ model light
curves arising from bootstrap fits in which the transit is 3$\sigma$
early, 1$\sigma$ early, and 1$\sigma$ late and 3$\sigma$ late. }
\end{figure}

\clearpage

\begin{deluxetable}{llll}
\tablecaption{Co-Planar Three-Planet $i=90^{\circ}$ Jacobi
Parameters of the Gl 876 System} \tablewidth{0pt} \tablehead{
Parameter$~~$ & Planet d$~~$ & Planet c$~~$ & Planet b$~~$ \\
} \startdata \tableline $m$\tablenotemark{a}      & 5.89 $\pm$ 0.54
$M_{\oplus}$ & 0.619 $\pm$
0.005 $M_{\rm Jup}$ & 1.935 $\pm$ 0.007 $M_{\rm Jup}$\\
$P$ (d)                   & 1.93776 $\pm$ 0.00007        & 30.340
$\pm$
0.013              & 60.940 $\pm$ 0.013\\
$K$ (m\,s$^{-1}$)         & 6.46 $\pm$ 0.59              & 88.36
$\pm$
0.72                & 212.60 $\pm$ 0.76\\
$a$\tablenotemark{a} (AU) & 0.0208067 $\pm$ 0.0000005    & 0.13030
$\pm$
0.00004           & 0.20783 $\pm$ 0.00003\\
$e$                       & 0 (fixed)                    & 0.2243
$\pm$
0.0013             & 0.0249 $\pm$ 0.0026\\
$\omega$ ($^{\circ}$)     & 0 (fixed)                    & 198.3
$\pm$
0.9                 & 175.7 $\pm$ 6.0\\
$M$ ($^{\circ}$)          & 309.5 $\pm$ 5.1              & 308.5
$\pm$
1.4                 & 175.5 $\pm$ 6.0\\
transit epoch             & JD 2452490.756 $\pm$ 0.027   & JD
2452517.633 $\pm$ 0.051 & \\
\enddata
\tablenotetext{a}{Quoted uncertainties in planetary masses and
semi-major axes {\it do not} incorporate the uncertainty in the mass
of the star} \label{3plparam}
\end{deluxetable}

\end{document}